\documentclass[twocolumn,showpacs,preprintnumbers,amsmath,amssymb]{revtex4}

\usepackage{graphicx}
\usepackage{dcolumn}
\usepackage{bm}

\begin{document}


\title{Superconducting Gap Function in Antiferromagnetic Heavy-Fermion UPd$_2$Al$_3$ Probed by Angle Resolved Magnetothermal Transport Measurements}

\author{T.~Watanabe$^1$, K.~Izawa$^1$,  Y.~Kasahara$^1$, Y.~Haga$^2$,Y.~Onuki$^{2,3}$, P.~Thalmeier$^4$, K.~Maki$^5$, and Y.~Matsuda$^1$}
\affiliation{$^1$Institute for Solid State Physics, University of Tokyo, Kashiwanoha 5-1-5, Kashiwa, Chiba 277-8581, Japan}%
\affiliation{$^2$Advanced Science Research Center, Japan Atomic Energy Research Institute, Tokai, Ibaraki 319-1195I, Japan }%
\affiliation{$^3$Graduate School of Science, Osaka University, Toyonaka, Osaka 560-0043, Japan}%
\affiliation{$^4$Max-Planck-Institute for the Chemical Physics of Solid, N\"othnitzer Str.40, 01187 Dresden, Germany}%
\affiliation{$^5$Department of Physics and Astronomy, University of Southern California, Los Angeles, CA 90089-0484}%

\date{\today}

\begin{abstract}

The superconducting gap structure of heavy fermion UPd$_2$Al$_3$, in which unconventional superconductivity coexists with antiferromagnetic (AF) order with atomic size local moments, was investigated by the thermal conductivity measurements in a magnetic field  {\boldmath $H$} rotating in various directions relative to the crystal axes.  The thermal conductivity displays distinct two-fold oscillation when  {\boldmath $H$} is rotated in the plane orthogonal to the basal $ab$-plane, while no oscillation was observed when  {\boldmath $H$} is rotated within the basal plane.    These results provide strong evidence that the gap function $\Delta$({\boldmath $k$}) has a single line node orthogonal to the $c$-axis located at the AF Brillouin zone boundary, while $\Delta$({\boldmath $k$}) is isotropic within the basal plane.  This gap structure indicates that the pairing interaction in neighboring planes strongly dominates over the interaction in the same plane.   The determined nodal structure is compatible with the resonance peak in the dynamical susceptibility observed in neutron inelastic scattering experiments.  Based on these results, we conclude that the superconducting pairing function of UPd$_2$Al$_3$ is most likely to be $d$-wave with a form $\Delta$({\boldmath $k$}$)=\Delta_0\cos(k_zc)$. 

\end{abstract}


\maketitle	

\section {Introduction}

	The relationship between superconductivity and magnetism is a fundamental problem in the physics of strongly correlated electron systems.  Over the past two decades, unconventional superconductivity with symmetries other than $s$-wave symmetry has been found in several classes of materials, including heavy fermion (HF), high-$T_c$, ruthenate, and organic superconductors.  Unconventional superconductivity is characterized by anisotropic superconducting gap functions belonging to a nontrivial representation of the crystal symmetry group, which may have zeros (nodes) along certain directions in the Brillouin zone \cite{sigrist,mineev}.  The nodal structure is closely related to the pairing interaction of the electrons and it is widely believed that the presence of nodes in almost all unconventional superconductors discovered up till now is a signature of a magnetically mediated interaction, instead of the conventional electron-phonon mediated interaction.   HF compounds containing Ce and U atoms are known to exhibit a rich variety of unconventional superconductivity associated with their peculiar magnetic properties \cite{thalmeier1}.  In these materials, as the temperature is lowered, $f$-electrons, which are well localized with well-defined magnetic moments at high temperatures, begin to become delocalized due to the hybridization of atomic and conduction electron wave functions.  Eventually, at very low temperatures, the $f$-electrons appear to become itinerant with  heavy effective electron mass about one hundred times the free electron mass.   In HF superconductors, the strong Coulomb repulsion within the atomic $f$-shells leads to a notable many-body effect and  often gives rise to Cooper pairing states with angular momentum greater than zero.   Therefore understanding the nodal structure of HF superconductors is important in understanding the physics of unconventional superconductivity in strongly correlated system. 

	Among HF superconductors, UPd$_2$Al$_3$ has aroused great interest because it displays quite unique properties.  In UPd$_2$Al$_3$, superconductivity with heavy mass (Sommerfelt specific coefficient $\gamma$=140~mJ/K$^2$mol) occurs at $T_c$=2.0~K after antiferromagnetic (AF) ordering with atomic size local moments ($\mu=$0.85$\mu_B$) occurs at $T_N$=14.3~K \cite{geibel}.   Below $T_c$, superconductivity coexists with magnetic ordering.  The ordered moments are coupled ferromagnetically and lie in the basal hexagonal $ab$-plane along the $a$-axis.  The ferromagnetic sheets are stacked antiferromagnetically along the $c$-axis with wave vector {\boldmath $Q_0$}=(0,0,$\pi/c$), where $c$ is the $c$-axis lattice constant \cite{kita} (For the structure of the hexagonal basal plane, see the inset of Fig.~1).   Due to a pronounced magnetic anisotropy, the aligned spins tend to fluctuate in the basal plane.   The presence of large local moments is in contrast to  other HF superconductors, in which static magnetic moments are absent or very small if present \cite{thalmeier1,uge}.   In UPd$_2$Al$_3$, both the superconductivity and AF ordering are associated with the 5$f$ electrons in the U atoms.  The presence of the two subsystems,  partly localized and partly itinerant, has been extensively studied in the context of the dual nature of strongly correlated electrons \cite{thalmeier1,machale,zwicknagl,sato}. 
			
	In the superconducting state of UPd$_2$Al$_3$,  two noticeable features have been reported.  The first is the modulation of the tunneling conductivity above the superconducting gap at an energy of about 1.2~meV in cross-type tunnel junctions UPd$_2$Al$_3$-AlO$_x$-Pb,  on a UPd$_2$Al$_3$ thin film \cite{jourdan}.  Furthermore,  neutron inelastic scattering experiments have revealed a strongly damped AF spin-wave excitation (magnetic exciton) with an excitation energy of about 1.5~meV at and in the vicinity of  {\boldmath $Q_0$} well below $T_N$.  Such an anomaly in the tunneling spectrum is well known from ordinary strong coupling electron-phonon superconductors like Pb.  In analogy, the ''strong coupling anomalies" observed in UPd$_2$Al$_3$  were attributed to the strong interaction between the HF quasiparticles and AF spin-wave excitations.   The second feature is the appearance of a ''resonance peak" in the neutron inelastic scattering in the vicinity of  {\boldmath $Q_0$} at about 0.3~meV well below $T_c$ \cite{metoki,bernhoeft1,sato}.    This peak has been interpreted as the creation of quasielectron/quasihole pairs from superconducting  condensed pairs by magnetic neutron scattering,  which in turn appears to be closely related to the sign of the superconducting order parameter under the transition   {\boldmath $k$}$\rightarrow$ {\boldmath $k+Q_0$} \cite{bernhoeft2}.   It should be noted that a resonance peak in the superconducting state has also been reported in $d$-wave high-$T_{c}$ cuprates \cite{fong} and borocarbide superconductors \cite{kawano} but its origin is controversial. 
	
	A  major outstanding question about UPd$_{2}$Al$_{3}$ is the nature of the microscopic pairing interaction responsible for the superconductivity.   To elucidate the pairing mechanism,  identification of the symmetry of the superconducting order parameter is of primary importance.    In particular, the gap structure of UPd$_2$Al$_3$ is expected to provide most valuable information on the relationship between the superconductivity and magnetism.  The gap function of UPd$_2$Al$_3$ has been the subject of extensive studies.  Thermodynamic  and nuclear magnetic resonance (NMR) measurements have revealed that the gap function is anisotropic \cite{caspary,tou}.  In particular, the NMR relaxation rate $T_1^{-1}$,  which exhibits a $T^3$ behavior in the superconducting state over four orders of magnitude down to 0.1$T_c$, along with the absence of the Hebel-Slichter coherence peak, indicate the existence of line nodes in the gap function \cite{tou}.   Pauli limiting $H_{c2}$ \cite{hessert}, NMR Knight shift reduction \cite{tou}, and $\mu$SR reduction in spin susceptibility \cite{mu} below $T_c$  favors a spin singlet state.  Thus spin-singlet superconducting symmetry with line nodes in UPd$_2$Al$_3$ seems to be confirmed.		
	
	 It has been pointed out  that the resonance peak around  {\boldmath $Q_0$} in the neutron inelastic scattering spectrum implies an order parameter displaying sign inversion on translation by  {\boldmath $Q_0$} on the Fermi surface, due to the coherence factor in the dynamical susceptibility of the conduction electrons.   In the simplest scenario,  $\Delta$({\boldmath $k$}) $= -\Delta$({\boldmath $k+Q_0$}) is inferred,  which suggests the existence of nodes orthogonal to the $c$-axis (horizontal node),   \cite{bernhoeft1,bernhoeft2,comment1}.  Here $\Delta$({\boldmath $k$}) is the superconducting order parameter.   Although the coherence factor argument may place a constraint on the sign of the order parameter,  the neutron scattering intensity mainly arises from a small minor band, which seems to be irrelevant in the occurrence of superconductivity, as suggested in Ref. \cite{bernhoeft2}.  Moreover, the neutron scattering results do not provide information on the position and number of the horizontal nodes and of the gap structure within the basal plane (vertical node).   The existence of  horizontal nodes was also suggested by the tunneling conductance along the $c$-axis \cite{jourdan}.  However the tunnelling conductance in anisotropic superconductors strongly depends on the tunneling process of the junction, whether the process is coherent or incoherent,  and on the transfer matrix originating from the atomic orbital at the surface, as extensively studied in high-$T_c$ cuprates \cite{xiang,gaifullin,hussey}.  In fact,  the typical $d$-wave gap structure has never been observed in $c$-axis tunneling experiments of high-$T_c$ cuprates.    On the other hand, the presence of  line nodes perpendicular to the basal plane is suggested by the anisotropy of the thermal conductivity \cite{chiao}.  Thus, despite great experimental effort, the detailed structure of the superconducting order parameter is still an unresolved issue.  therefore, there is a need for a probe of the quasiparticle excitation with spatial resolution in  {\boldmath $k$}-space.  
	 	
	The purpose of this work is to determine the superconducting order parameter $\Delta(${\boldmath $k$}) of UPd$_2$Al$_3$ by measuring the angle resolved magneto-thermal transport, which has proven to be a powerful probe for determining low energy quasiparticle excitations,  including  direction \cite{vekhter,vekhter3,maki,nakai,yu,aubin,ocana,izawaRu,izawaCe,aoki,izawabedt,izawaboro,park,izawaPr,deguchi}.  The thermal conductivity is a unique transport quantity that does not vanish in the superconducting state, responding to unpaired quasiparticles that carry the heat.    We provide  direct evidence that the gap function has a horizontal node located at the AF zone boundary $k_z=\pm \pi/2c$, and is isotropic within the basal plane.   These results place strong constraints on models that attempt to explain the mechanism of the unconventional superconductivity of UPd$_2$Al$_3$.   We discuss the pairing mechanism infered by the gap structure.		
	
\section {Experimental}

	High quality single crystals of UPd$_2$Al$_3$ with $T_c=2.0$~K were grown by the Czochralski pulling method in a tetra-arc furnace.    The residual resistivity ratio was 55 along the $b$-axis and 40 along the $c$-axis, indicating the crystals were of the highest crystal quality currently achievable.  The upper critical fields in {\boldmath $H$} parallel to the $a$-,  $b-$, and $c$-axes at 0.4~K, determined by the resistive transitions,  were $H_{c2}^a$=3.2~T ,$H_{c2}^b$=3.1~T, and $H_{c2}^c$=3.8~T, respectively.   We measured thermal conductivity along the $c$-axis of the hexagonal crystal structure, $\kappa_{zz}$ (heat current {\boldmath $q$} $\parallel$ $c$) and along the $b$-axis $\kappa_{yy}$ ({\textit {\textbf q}} $\parallel$ $b$) by the steady-state method.   The sample dimensions for the $\kappa_{zz}$ and $\kappa_{yy}$ measurements were 0.32$\times$0.34$\times$2.50 mm$^3$ and 0.22$\times$0.50$\times$1.80 mm$^3$,  respectively.   To apply {\boldmath $H$} with high accuracy relative to the crystal axes, we used a system with two superconducting magnets generating magnetic fields in two mutually orthogonal directions and a $^3$He cryostat set on a mechanical rotating stage at the top of a Dewar.  By computer-controlling the two superconducting magnets and rotating stage, we were able to rotate  {\boldmath $H$} with a misalignment of less than 0.5$^{\circ}$ from each axis, which we confirmed by the simultaneous measurements of the resistivity.   
	
	In the present experiments, $\kappa_{yy}$ and $\kappa_{zz}$ were measured  keeping {\boldmath $H$} always perpendicular to  {\boldmath $q$},  i.e. the heat current perpendicular to the vortices dominates.  This configuration ({\boldmath $H$}$\perp${\boldmath $q$}) is important because when the angle between {\boldmath $q$} and {\boldmath $H$} is changed by rotating {\boldmath $H$}, an additional angular variation of the thermal conductivity appears as a function of the angle between {\boldmath $H$} and {\boldmath $q$} due to the difference in the effective density of states (DOS) for quasiparticles traveling parallel to the vortices and for those moving in the perpendicular direction.   In the field rotation experiments, the thermal conductivity was measured after field cooling above $T_{c}$.  Consecutive measurements with an inverted direction produced little hysteresis in the angular dependence of the thermal conductivity, which indicates that field trapping related to vortex pinning was negligible. 
	
\begin{figure}[t]
\includegraphics[scale=0.7,angle=0]{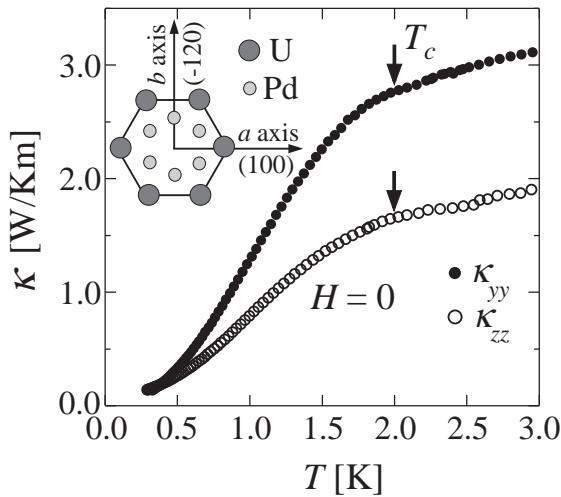}
\caption{Temperature dependence of the thermal conductivity in zero field in the $b$-axis $\kappa_{yy}$ ({\boldmath $q$}$\parallel b$) and  $c$ axis $\kappa_{zz}$ ({\boldmath $q$}$\parallel c$).  Inset shows the structure of the hexagonal basal plane of UPd$_{2}$Al$_{3}$ with the alignment of the $a$-axis (100) and $b$-axis (-1,2,0), as used in the text.   }
\end{figure}

	Figure 1 depicts the temperature dependence of $\kappa_{yy}$ and $\kappa_{zz}$ in zero field.  The magnitude and $T$-dependence of $\kappa_{yy}$ and $\kappa_{zz}$ are similar to those reported in Ref. \cite{chiao,hiroi}.    In UPd$_2$Al$_3$ there are three contributions that carry the heat; electrons, phonons and spin-waves.  The spin-wave contribution appears to be negligible below $T_c$, because a spin-wave has a finite gap of $\sim1.5$~meV at the zone center and cannot contribute to the heat conduction at low temperatures.  The Wiedemann-Franz ratio $L=\frac{\kappa}{T} \rho$ at $T_c$ is 0.95$L_0$ for $\kappa_{zz}$ and is 1.16$L_{0}$ for $\kappa_{yy}$, where $L_{0}$ is the Lorentz number and $\rho$ is the resistivity .  These results indicate that the electron contribution strongly dominates over the phonon contribution,  at least below $T_c$.  Unfortunately the present temperature range (the lowest temperature was 0.28~K) is not low enough to determine the detailed $T$-dependence and residual values of $\kappa_{yy}$ and $\kappa_{zz}$, which would provide important information about the gap structure.  We therefore do not discuss these subjects in this paper.
	
	Figure 2 shows the $H$-dependence of $\kappa_{yy}$ for {\boldmath $H$}$\parallel a$ and {\boldmath $H$}$\parallel c$ below $T_c$.  In both field directions, $\kappa_{yy}$ increases with $H$ after an initial decrease at low fields.  For  {\boldmath $H$}$\parallel c$, $\kappa_{yy}$ increases almost linearly with $H$, $\kappa_{yy}\propto H$, at 0.36~K.  The consequent minimum is much less pronounced at lower temperatures.   As $H$ approaches $H_{c2}^a$ or $H_{c2}^c$,  $\kappa_{yy}$ shows a steep increase and attains its normal state value.  In the normal state above $H_{c2}$, $\kappa_{yy}$ shows a slight decrease, which is scaled by the positive magnetoresisitance, supporting the conclusion that the electron contribution dominates the thermal conductivity.   
	
\begin{figure}[t]
\includegraphics[scale=0.7]{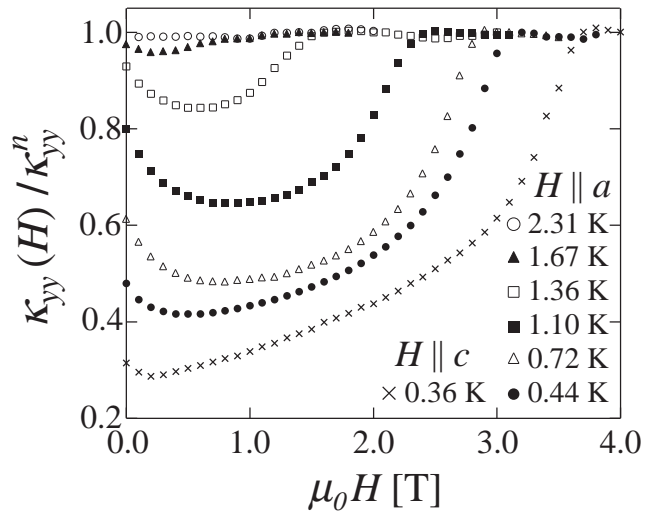}
\caption{Magnetic field dependence of the $b$-axis thermal conductivity $\kappa_{yy}$ for {\boldmath $H$}$\parallel a$ and {\boldmath $H$}$\parallel c$. $\kappa_{yy}$ is normalized by the normal state value just above the upper critical field $\kappa_{yy}^{n}$.   }
\end{figure}
	
	The $H$-dependence of the thermal conductivity in the superconducting state is markedly different from that observed in ordinary $s$-wave superconductors,  in which the thermal conductivity shows an exponential behavior with  much slower growth with $H$ at $H \ll H_{c2}$ \cite{boakin}.  The understanding of the heat transport for superconductors with nodes has progressed considerably during past few years \cite{barash,kubert,vekhter2}.  In contrast to classical superconductors, the heat transport in nodal superconductors is dominated by contributions from delocalized quasiparticle states rather than bound states associated with vortex cores.  The most remarkable effect on the thermal transport is the Doppler shift of the energy of quasiparticles with momentum {\boldmath $p$} in the circulating supercurrent flow $\mbox{\boldmath $v$}_s$  ($E(\mbox{\boldmath $p$})\rightarrow E(\mbox{\boldmath $p$})-\mbox{\boldmath $v$}_s \cdot \mbox{\boldmath $p$}$)  \cite{volovik,hussey,nakai2}.  This effect (Volovik effect) becomes important at positions where the local energy gap becomes smaller than the Doppler shift term ($\Delta < \mbox{\boldmath $v$}_s \cdot \mbox{\boldmath $p$}$), which can be realized in the case of superconductors with nodes.  In the presence of line nodes where the density of states (DOS) of electrons $N(E)$ has a linear energy dependence ($N(E)\propto E$), $N(H)$ increases in proportion to $\sqrt{H}$.  At high temperatures and low fields, where the condition $\sqrt{H/H_{c2}}<T/T_{c}$ is satisfied, the thermally excited quasiparticles dominate over the Doppler shifted quasiparticles.  It has been shown that in this regime, while the Doppler shift enhances the DOS, it also leads to a suppression of both the impurity scattering time and Andreev scattering time off the vortices \cite{kubert,vekhter2,izawaRu,izawaCe,won}.  This suppression can exceed the parallel rise in $N(E)$ at high temperatures and low fields, which results in nonmonotonic field dependence of the thermal conductivity.  As seen in other superconductors with nodes, the initial decrease of the thermal conductivity disappears at low temperatures.   The steep increase of $\kappa_{yy}$ near $H_{c2}$, which is also observed in pure Nb, UPt$_3$, and Sr$_2$RuO$_4$, may be due to the enhancement of the quasiparticle mean free path caused by tunneling between vortex cores, which is possible near $H_{c2}$.    Thus the $H$-dependence of $\kappa_{yy}$ in UPd$_2$Al$_3$,  initial decrease at low field at high temperatures and linear behavior $\kappa_{yy}\propto H$ at low temperatures,  are in agreement with the existence of line nodes in $\Delta(\mbox{\boldmath $H$})$.  We note that quite similar $H$-dependence has been reported in Sr$_2$RuO$_4$ with an anisotropic gap function \cite{izawaRu}.
	
\begin{figure}[b]
\includegraphics[scale=0.5]{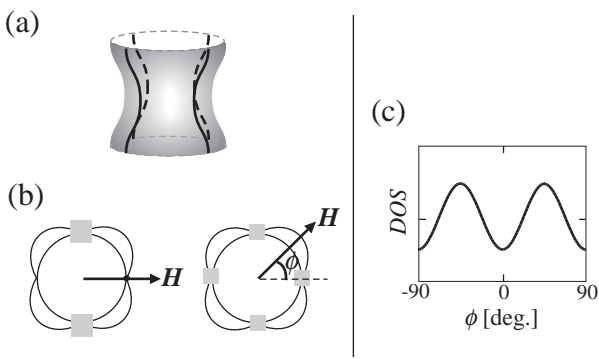}
\caption{(a)Schematic figure of the gap structure with four line nodes perpendicular to the basal plane (vertical node). (b)Schematic diagram showing the regions on the Fermi surface that experience the Doppler shift in {\boldmath $H$} within the basal plane.  We have assumed $d_{xy}$ symmetry.  $\phi$ = ({\boldmath $H$},$a$) is the azimuthal angle measured from the $a$-axis.   With {\boldmath $H$} applied along the antinodal directions, all four nodes contribute to the DOS, while for {\boldmath $H$} applied parallel to the node directions, the Doppler shift vanishes at two of the nodes. (c) Four-fold oscillation of the DOS for {\boldmath $H$} rotating in the basal plane.  The DOS shows a maximum (minimum) when {\boldmath $H$} is applied in the anitinodal (nodal) direction. }
\end{figure}
		
	We here briefly discuss the band structure of  UPd$_2$Al$_3$.   According to band calculations and de Haas-van Alphen measurements in the AF phase, the Fermi surface has four principle sheets of predominantly 5$f$-electron character, which are labeled  "cigar", "eggs","cylinder", and "party hat" \cite{zwicknagl,knopfle,inada}.    The largest Fermi sheet with large electron mass and the strongest 5f-admixture is the "cylinder", which has the shape of a corrugated cylinder with hexagonal in-plane anisotropy.   Therefore it is natural to consider that the thermal conductivity is mainly determined by this Fermi sheet.    In what follows, we assume that the cylindrical Fermi sheet with heavy mass is responsible for the superconductivity, neglecting all other sheets.

\section{Angular-dependent magnetothermal conductivity}

\begin{figure}[b]
\includegraphics[scale=0.5]{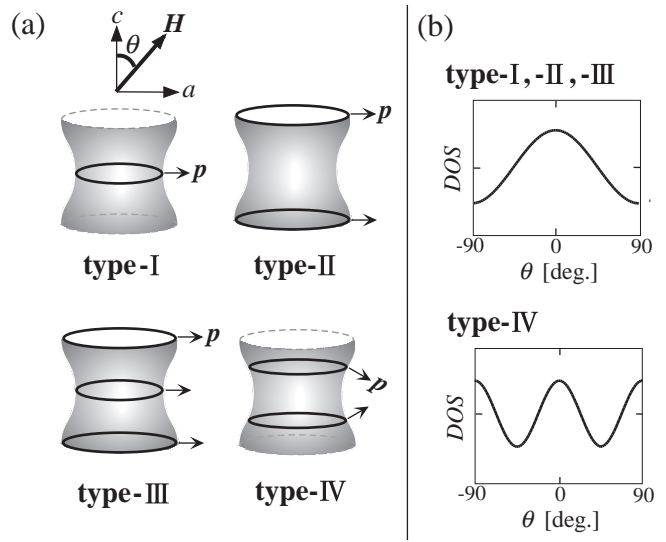}
\caption{ (a) Schematic figure of the gap structure with line nodes parallel to the basal plane (horizontal node).  Line nodes are located at the bottleneck (type-I) and at the AF zone boundary (type-II).  Two line nodes are located at the bottleneck and the AF zone boundary (type-III) and at positions shifted off the bottleneck (type-IV) .   (b) Oscillations of the DOS for {\boldmath $H$} rotating in the $ac$-plane for various gap functions.  Two-fold oscillation with the same sign are expected for type-I, -II, and -III.  On the other hand, for type-IV, an oscillation with a double minimum is expected.}
\end{figure}

	Having established the predominant contribution of the delocalized quasiparticles in the thermal transport, the next issue is the nodal structure of UPd$_2$Al$_3$.  An important advantage of measuring the thermal conductivity is that it is a directional probe of the nodal structure.   Recently measurement of the thermal conductivity or heat capacity with {\boldmath $H$} applied various directions relative to the crystal axes has been established as a powerful method to determine the superconducting gap structures in {\boldmath $k$}-space.     At the heart of this method is the Volovik effect discussed previously.  Since quasiparticles contribute to the DOS when the Doppler-shifted energy exceeds the local energy gap  ($\Delta(\mbox{\boldmath $k$}) < \mbox{\boldmath $v$}_s \cdot \mbox{\boldmath $p$}$),  an immediate conclusion is that the DOS depends sensitively on the angle between {\boldmath $H$}  and the direction of the nodes.  The periodicity, phase, and shape of typical oscillations in the angular variation of the thermal conductivity give direct information of the gap structure, including the type of nodes (point or line) and their direction.  Therefore, possible allowed gap functions can be much more restricted in {\boldmath $k$}-space.  It should be noted that  this method does not rely on any specific mechanism of the superconductivity.			
	
	We first discuss the case when {\boldmath $H$} is rotated within the basal $ab$-plane.  Here, for simplicity, we assume $d_{xy}$ symmetry with four line nodes located perpendicular to the basal plane, and assume a circular cross section of the Fermi sheet, neglecting the hexagonal in-plane anisotropy, as illustrated in Figs.~3 (a) and (b).    Here $\phi$=({\boldmath $H$},$a$) is the azimuthal angle measured from the $a$-axis.  One expects a significant anisotropy between the field induced DOS for {\boldmath $H$} parallel to the nodal directions and {\boldmath $H$} parallel to the anti-nodal direction.  The Doppler effect should also lead to angular dependence of the DOS for {\boldmath $H$} rotated within the basal plane.  A minimum in the DOS occurs when {\boldmath $H$} is along the nodal directions.  Since, in that case, at two of the four nodes,  the circulating currents are in the plane orthogonal to the momentum vector at the nodes, the Doppler shift vanishes; only two nodes contribute to the DOS, as illustrated in Fig.~3.  In contrast, when {\boldmath $H$} is directed along the antinodal directions, the Doppler shift is non-zero at some points in space for each node and all four nodes contribute to the DOS.  As a result, the DOS oscillates with four-fold symmetry, as shown in Fig.~3(c).  The amplitude of the DOS oscillation $\delta N(E)/N(E)$ appears to be model dependent, but all calculations predict an amplitude ranging from 3\% to 10\% at zero temperature.   In fact, a clear four-fold modulation of the thermal conductivity, which reflects the angular position of the nodes, has been observed in the high-$T_c$ cuprate YBa$_2$Cu$_3$O$_{7-\delta}$ \cite{yu,aubin,ocana}, heavy fermion CeCoIn$_5$ \cite{izawaCe,aoki}, and organic $\kappa$-(BEDT-TTF)$_2$Cu(NCS)$_2$ \cite{izawabedt} with $d$-wave symmetries and in borocarbide YNi$_{2}$B$_{2}$C with $s+g$-wave symmety \cite{izawaboro,watanabe},  while such an oscillation is absent in the fully gapped Y(Ni$_{0.95}$Pt$_{0.05}$)$_{2}$B$_{2}$C with $s$-wave symmetry \cite{kamata}, demonstrating that  thermal conductivity can be a relevant probe of the superconducting gap structure.  In general,  $\kappa_{zz}$ oscillates with $n$-fold symmetry corresponding to the number of vertical nodes $n$.  
	
	We next discuss the situation when {\boldmath $H$} is rotated within the $ac$-plane perpendicular to the basal $ab$-plane.    The Fermi surface has open orbits along the $c$-axis.  We here consider four gap functions in the AF Brillouin zone shown in Fig.~4(a).  
\begin{enumerate}
\item type-I : A horizontal node located at the bottleneck;   $\Delta(\mbox{\boldmath $k$}) \propto \sin k_zc$. 
\item type-II : A horizontal node located at the AF zone boundary;  $\Delta(\mbox{\boldmath $k$}) \propto \cos k_zc$.
\item type-III: A hybrid of type-I and -II. Two horizontal nodes located  at the bottleneck and the AF zone boundary; $\Delta(\mbox{\boldmath $k$}) \propto \sin 2k_zc$.
\item type-IV : Two horizontal nodes located at  positions shifted off the bottleneck in the Brillouin zone;  $\Delta(\mbox{\boldmath $k$}) \propto \cos 2k_zc$.
\end{enumerate}
The expected angular variations corresponding to these gap functions are shown schematically in Fig.~4(b).   Here $\theta$ is the polar angle between {\boldmath $H$} and the $c$-axis.  The angular variation of the Doppler shifted DOS is a function of the relative angle between {\boldmath $H$} and {\boldmath $p$}.  Therefore the twofold oscillations with the same phase are expected for type-I, -II , and -III gap functions, in which the horizontal nodes are located at the position where {\boldmath $p$}$\parallel ab$-plane; one cannot distinguish these three gap functions when the Fermi surface has an open orbit along the $c$-axis.  For type-IV, one expects an oscillation with a double maximum structure as a function of $\theta$.  	
	
	We note that the angular variation is dominated by the Doppler shifted DOS at low fields.   On the other hand, at high fields near $H_{c2}$, there are additional contributions which also cause the oscillation of the DOS,  i.e. the anisotropy of the Fermi velocity and $H_{c2}$ \cite{nakai,udagawa}.  Generally, the oscillation due to the Fermi velocity and due to $H_{c2}$ have the same sign.  Therefore,  it is desirable to determine the nodal structure at low fields where the oscillation arising from $H_{c2}$ and the Fermi velocity is negligible.

\section {Results and disscussion}

\subsection{Vertical nodes}

\begin{figure}[t]
\includegraphics[scale=0.6]{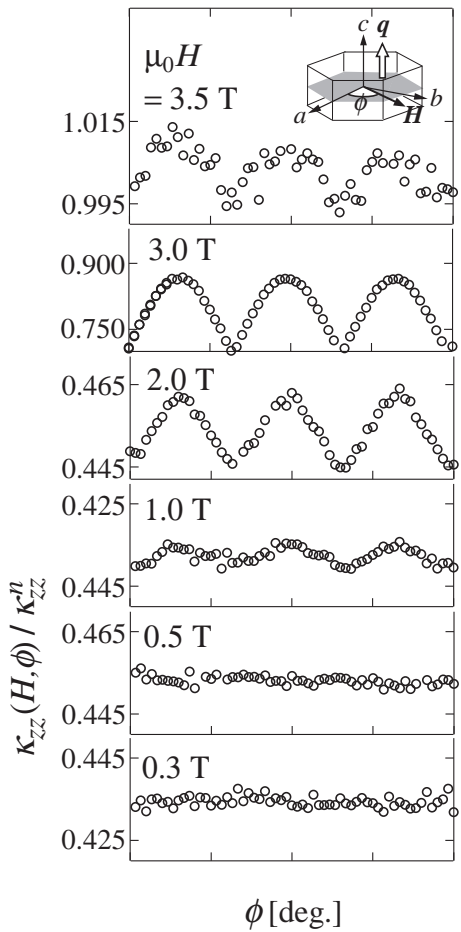}
\caption{Angular variation of the $c$-axis thermal conductivity  $\kappa_{zz}$({\boldmath $H$},$\phi$) normalized by the normal state value, $\kappa_{zz}^n$, at several fields at 0.4~K. {\boldmath $H$} was rotated within the basal plane (see  inset).  A distinct six-fold oscillation is observed above 0.5~T, while oscillation is absent at 0.5 and 0.3~T.  At $H$=3.5~T ($>H_{c2}$), the system is in the normal state.   }
\end{figure}
	
	We examine here the possibility of vertical line nodes perpendicular to the basal plane.  Figure 5 shows $\kappa_{zz}$({\boldmath $H$},$\phi$) as a function of $\phi$ at 0.4~K,  measured by rotating {\boldmath $H$} within the basal plane.   Above 0.5~T, a distinct six-fold oscillation is observed in $\kappa_{zz}$({\boldmath $H$},$\phi$), reflecting the hexagonal symmetry of the crystal.  However, no discernible six-fold oscillation was observed below 0.5~T within our experimental resolution.   $\kappa_{zz}$ can be decomposed as $\kappa_{zz}=\kappa_{zz}^0+\kappa_{zz}^{6\phi}$ where $\kappa_{zz}^0$ is a $\phi$-independent term and $\kappa_{zz}^{6\phi}=C_{zz}^{6\phi}\cos{6\phi}$ has six-fold symmetry with respect to $\phi$ -rotation.   The sign of $C_{zz}^{6\phi}$ is negative in the whole $H$ range.  The six-fold oscillation is clearly observed even in the normal state above $H_{c2}$.   In contrast, the amplitude of the sixfold oscillation,  $|C_{zz}^{6\phi}|/\kappa_{zz}^n$,  is less than 0.2\% below 0.5~T, where $\kappa_{zz}^n$ is the $c$-axis thermal conductivity in the normal state just above $H_{c2}$.     In Fig.~6, the $H$-dependence of  $|C_{zz}^{6\phi}|/\kappa_{zz}^n$ is plotted as a function of $H$.     In the region $H_{c2}^{a}<H<H_{c2}^{b}$, $|C_{zz}^{6\phi}|/\kappa_{zz}^n$ is  largest.  This is because the sample is either in the normal or superconducting state with rotating  {\boldmath $H$}.   With decreasing $H$, $|C_{zz}^{6\phi}|/\kappa_{zz}^n$ decreases and vanishes below 0.5~T.    
	
	We point out here that the AF magnetic domain structure is responsible for the six-fold symmetry and the nodal structure is not related to the oscillation.   For UPd$_2$Al$_3$,  there are several AF ordered phases; three with the applied field in the easy (basal) plane, the (i)-,(ii)- , and (iii)-phases, and one along the magnetically hard $c$-axis \cite{kita}.   The magnetic structures in the (i)- and (ii)-phases are shown schematically in the inset of Fig.~6.  The transitions, (i)-(ii) and (ii)-(iii),  occur at 0.6 and 4.2~T, respectively, well below $T_N$.  The fact that the sixfold oscillation disappears in the vicinity the (i)-(ii) boundary strongly indicates that the six-fold oscillation is closely related to the AF magnetic structure.     This idea is reinforced by the fact that six-fold oscillation is observable even above $H_{c2}$.   In the (i)-phase, the ordered moments point to the $a$-axis, forming domain structures with six-fold degeneracy inherent to the hexagonal crystal structure.  The spin structure is not effected by {\boldmath $H$}-rotation in the basal plane.  On the other hand, for the (ii)-phase,  {\boldmath $H$}-rotation causes  domain reorientation, giving rise to  rotation of the ordered magnetic moments against a small anisotropy in the basal plane.  As a result, the magnetic domain structure changes with six-fold symmetry with {\boldmath $H$} rotation in the basal plane.   Since the quasiparticles are scattered at the domain boundaries, $\kappa_{zz}$ oscillates with six-fold symmetry in the (ii)-phase.  Thus the six-fold symmetry observed in $\kappa_{zz}$ above 0.5~T is most likely to be due to the magnetic domain structure.  In addition,  the in-plane anisotropy of $H_{c2}$ and the Fermi velocity is also important near $H_{c2}$.   The most important finding is the absence of the oscillation in the (i)-phase where the spin structure is independent of {\boldmath $H$} rotation.  This definitely indicates that {\it there are no nodes located perpendicular to the basal plane, i.e. gap function in the basal plane is isotropic.}    
	
\begin{figure}[t]
\includegraphics[scale=0.6]{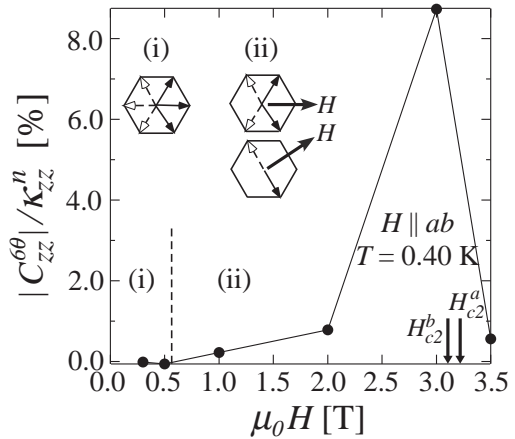}
\caption{Field dependence of the amplitude of the six-fold oscillation $C_{zz}^{6\phi}$ normalized by the normal state value,  $\kappa_{zz}^n$.  $H_{c2}^a=3.2$~T and $H_{c2}^b$=3.12~T are the upper critical fields in {\boldmath $H$} parallel to the $a$- and $b$-axis, respectively.  The inset shows the magnetic domain structures in parallel fields. The hexagon shows the basal plane.  Populated magnetic domains with non-parallel spins are indicated by filled arrows.  The open arrows indicate the antiparallel spins in the neighboring plane.  The transition (i)-(ii) occurs at about 0.6~T.   In the (i)-phase, the ordered moments point to the $a$-axis, forming domain structures with six-fold degeneracy.  In the (ii)-phase, the domain structure changes to six-fold symmetry with respect to the {\boldmath $H$}-rotation in the basal plane.}
\end{figure}
	
\subsection{Horizontal nodes}

	We next examine the horizontal line nodes parallel to the basal plane.  Figure 7 displays the angular variation of $\kappa_{yy}$({\boldmath $H$},$\theta$) for rotating  {\boldmath $H$} as a function of $\theta$ within the $ac$-plane at 0.4~K.  A distinct oscillation with two-fold symmetry is observed in the superconducting state.  In contrast to $\kappa_{zz}$, no discernible twofold oscillation was observed in the normal state above $H_{c2}$.   $\kappa_{yy}$({\boldmath $H$},$\theta$) can be decomposed as $\kappa_{yy}=\kappa_{yy}^0+\kappa_{yy}^{2\theta}$ where $\kappa_{yy}^0$ is a $\theta$-independent term and $\kappa_{yy}^{2\theta}=C_{yy}^{2\theta}\cos{2\theta}$ is a term with twofold symmetry with respect to $\theta$-rotation.  Figure~8(a) depicts the $H$-dependence of $C_{yy}^{2\theta}$ at 0.4~K.  For comparison, the $H$-dependence of $\kappa_{yy}$ for  {\boldmath $H$}$\parallel c$ at $T$=0.36~K is plotted in Fig.~8(b).  There are three regions denoted  (I), (II) and (III), below $H_{c2}$.  In the vicinity of $H_{c2}$ ((III)-region), where $\kappa_{yy}$ increases steeply with $H$, the sign of $C_{yy}^{2\theta}$ is negative and the amplitude $|C_{yy}^{2\theta}|/\kappa_{yy}^n$ is of the order of 10\%.  Here $\kappa_{yy}^n$ is  $\kappa_{yy}$ in the normal state just above $H_{c2}$.   With decreasing $H$, $C_{yy}^{2\theta}$ changes  sign at about 2.3~T and become positive in the region where $\kappa_{yy}$ for {\boldmath $H$}$\parallel c$ shows a linear $H$-dependence ((II)-region).     Below about 0.25~T, where the second sign change takes place,  $\kappa_{yy}$ decreases with $H$ ((I)-region).  In this region, $C_{yy}^{2\theta}$ becomes negative.    

\begin{figure}[t]
\includegraphics[scale=0.6]{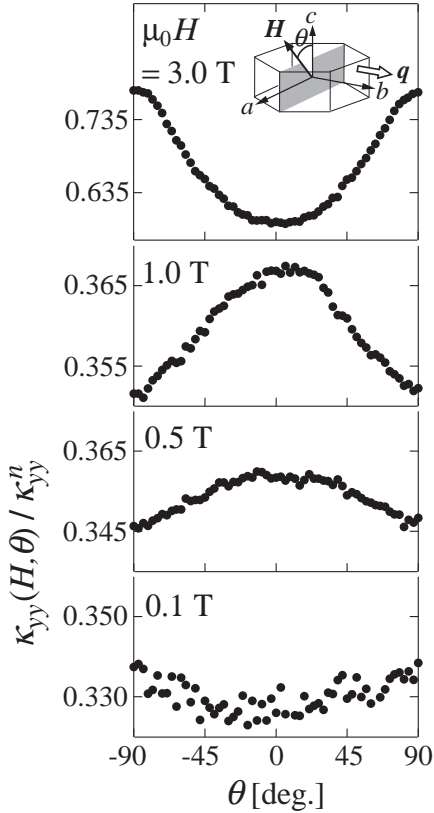}
\caption{Angular variation of the $b$-axis thermal conductivity  $\kappa_{yy}$({\boldmath $H$},$\theta$) normalized by the normal state value $\kappa_{yy}^n$ at several fields at 0.4~K.   {\boldmath $H$} was rotated within the $ac$-plane perpendicular to the basal plane (see inset). }
\end{figure}
	
	We here address the origins of the observed two-fold oscillation.  The disappearance of the oscillation above $H_{c2}$, together with the fact that there is only one magnetic phase in this configuration \cite{kita}, completely rule out the possibility that the origin is due to the magnetic domain structure.   There are two possible origins for the oscillation; the nodal structure and the anisotropy of the Fermi velocity and $H_{c2}$.  Obviously, as discussed previously, a large two-fold oscillation with negative sign observed in the (III)-region arises from the anisotropies of the Fermi velocity and  $H_{c2}$.   This immediately indicates that {\it the two-fold symmetry with positive sign in the (II)-region originates not from these anisotropies but from the quasiparticle structure associated with the nodal gap function.}  In addition, the amplitude of $C_{yy}^{2\theta}/\kappa_{yy}$ in the (II)-region is a few percent, which is quantitatively consistent with the prediction considering the Doppler shifted DOS.   We also note that the second sign change at low fields in the (I)-region is compatible with the nodal structure.  In this region, as discussed previously, the $H$-dependence of the thermal conductivity is governed by the suppression of the quasiparticle scattering rate.  As discussed in Ref. \cite{yu,aubin,ocana,izawaCe,vekhter3,maki,won},  the oscillation arising from the anisotropic carrier scattering time associated with the nodal structure also gives rise to the oscillation of $\kappa_{yy}$({\boldmath $H$},$\theta$) as a function of $\theta$.  In this case the sign of the oscillation is opposite to that arising form the Doppler shifted DOS in the (II)-region.   These considerations lead us to conclude that  UPd$_2$Al$_3$ has horizontal nodes.  In addition, the fact that there is a single maximum structure in the angular variation of $\kappa_{yy}$({\boldmath $H$},$\theta$) indicates that horizontal line nodes are located at  positions where the condition {\boldmath $p$}$\parallel ab$ in the  Brillouin zone is satisfied.  Thus {\it the allowed positions of the horizontal nodes are restricted at the bottleneck and  AF zone boundary}. 	
	
\begin{figure}[t]
\includegraphics[scale=0.55]{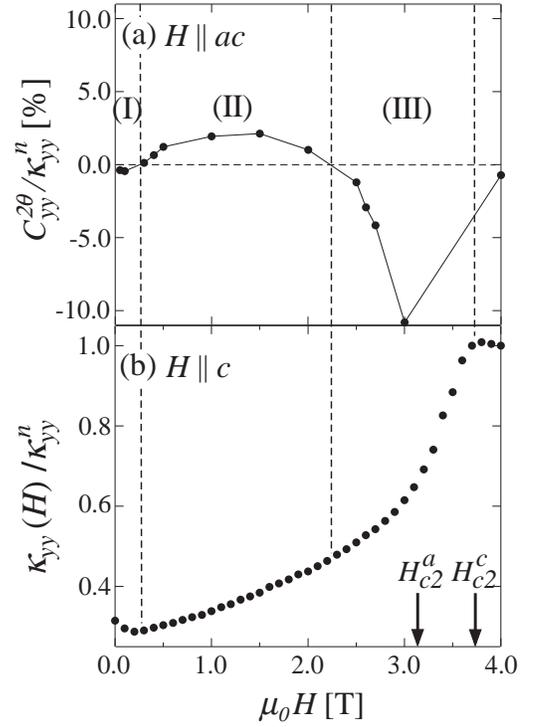}
\caption{(a) $H$-dependence of the amplitude of the two-fold symmetry $C_{yy}^{2\theta}$ normalized by the normal state thermal conductivity $\kappa_{yy}^n$ at 0.4~K.  The sign of $C_{yy}^{2\theta}$ is negative in the (I)- and (III)-regions, and is positive in the (II)-region.  (b) Field dependence of the $b$-axis thermal conductivity $\kappa_{yy}$ for {\boldmath $H$}$\parallel c$ at 0.36~K.  }
\end{figure}

\section{Conclusion}

\begin{figure}[t]
\includegraphics[scale=0.7]{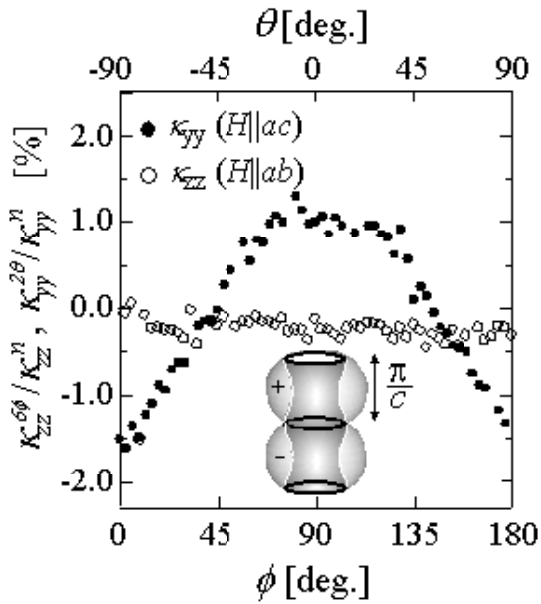}
\caption{Angular variation of the $b$-axis thermal conductivity $\kappa_{yy}^{2\theta}/\kappa_{yy}^{n}$ with rotating {\boldmath $H$} within the $ac$-plane and of the $c$-axis thermal conductivity $\kappa_{zz}^{6\phi}/\kappa_{zz}^{n}$ with rotating {\boldmath $H$} within the basal $ab$-plane at $T$=0.4~K and at $H$=0.5~T.  The amplitude of the six-fold oscillation in $\kappa_{zz}^{6\phi}/\kappa_{zz}^{n}$ is less than 0.2\% if it exists.  Inset: Schematic figure of the gap function of UPd$_{2}$Al$_{3}$ determined by angle resolved magnetothermal transport measurements. The thick solid lines indicate horizontal nodes located at the AF zone boundaries.    }
\end{figure}
	
	For comparison,  the angular variations of $\kappa_{yy}$ and $\kappa_{zz}$ at low fields are shown in Fig.~9.   While the amplitude of the two-fold oscillation $C_{yy}^{2\theta}/\kappa_{yy}$ is 3\%, which is quantitatively consistent with the Doppler shifted DOS, the amplitude of the six-fold oscillation $C_{zz}^{6\theta}/\kappa_{zz}$ is less than 0.2\%, which is more than 20 times smaller than the oscillation expected from the Doppler shifted DOS in the presence of nodes.    Combining the results, we arrive at the conclusion that {\it the gap function is isotropic in the basal plane and has horizontal node.}   The order parameters allowed,  by  thermal conductivity measurements,  are, 
\begin{enumerate}
\renewcommand{\labelenumiii}{(\roman{enumi})}
\item $\Delta(\mbox{\boldmath $k$})=\Delta_0 \sin k_zc$, 
\item $\Delta(\mbox{\boldmath $k$})=\Delta_0 \sin 2k_zc$   ~~~and 
\item $\Delta(\mbox{\boldmath $k$})=\Delta_0 \cos k_zc$. 
\end{enumerate} 
which are shown in the type-I, -II and -III gap structures in Fig.~4(a).  The first and second represent spin triplet gap functions, and the third represents a spin singlet gap function.  As discussed previously,  the $H_{c2}$, NMR, and $\mu$SR experiments all favor spin singlet pairing.  We note  that  spin singlet pairing is also supported by tunneling spectroscopy measurements.  In anisotropic superconductors, zero-bias conductance peak appears due to mid gap Andreev states  where the injected and reflected electrons feel the different signs of the pair potentials \cite{TK}.   In the experiment reported in Ref.~\cite{jourdan}, in which  tunneling conductance is measured along the $c$-axis, a zero-bias conductance peak appears for triplet paring while it is absent for singlet pairing.  The absence of the zero-bias conductance peak supports  singlet pairing.  Furthermore,  the third gap function is compatible with the constraint implied by the neutron resonance peak,  $\Delta$({\boldmath $k$}) $= -\Delta$({\boldmath $k+Q_0$}).   These considerations lead us to conclude that {\it the gap function of UPd$_2$Al$_3$ is most likely to be $\Delta(\mbox{\boldmath $k$})=\Delta_0 \cos k_zc$ with $d$-wave symmetry}, shown in the inset of Fig.~9.     

	Theoretical models with various nodal structures for UPd$_2$Al$_3$ have been proposed.  The determined gap structure cannot be explained in the framework of the electron-phonon coupling mechanism.    There are two classes of theories of the mechanism of the unconventional superconductivity in UPd$_2$Al$_3$.  The first is  superconductivity mediated by the exchange of  AF spin-wave excitation (magnetic excitons); the coupling between localized- and itinerant-subsystem is essential for the occurrence of superconductivity \cite{thalmeier1,machale,sato}.   The second is a mechanism based on the AF spin fluctuation arising from the strong electron-electron correlation in the itinerant subsystem, ignoring the localized subsystem \cite{nishikawa}.  This model predicts vertical nodes, which is inconsistent with the present results.   The existence of  horizontal nodes strongly supports the interpretation inferred from the resonance peak observed at the AF Bragg point  in the neutron inelastic scattering experiments.   The horizontal node located at the AF zone center indicates that pair partners cannot reside in the same basal plane.   The interlayer pairing appears to indicate that strong dispersion of the magnetic excitation along $k_{z}$ causes the pairing, as suggested in the magnetic exciton mediated superconductivity model \cite{thalmeier1,machale,sato}.  This dispersion is connected with the fluctuation of the out-of-plane component of the ordered moments, because the in-plane moments are saturated at $T_{c}\ll T_{N}$ and should not have large fluctuations.  The isotropic gap function in the basal plane implies that  the pairing interaction in the neighboring planes strongly dominates over the interaction in the same plane. This is incompatible with the theory in which only the electron correlation in the basal plane is responsible for the superconductivity.    Although the pairing interaction inferred from the determined gap function should be further scrutinized,  the recent results imply that the interlayer pairing interaction associated with the  AF interaction is most likely to be the origin of the unconventional superconductivity in UPd$_2$Al$_3$.  
	  
\section{Acknowledgements}

We thank H.~Ikeda, M.~Kofu, K.~Machida, K.~Miyake, N.~Metoki,  N.~K.~Sato, and I.~Vekhter,  for helpful discussions.  This work was partly supported by a Grant-in-Aid for Scientific Research Priority Area "Skutterudite" (No.15072206) of the Ministry of Education, Culture, Sports, Science and Technology, Japan.


\begin{thebibliography}{99}
\bibitem{uge}UGe$_{2}$ has a strong ferromagnetic moment in part of the superconducting regime, but is not really a HF material.
\bibitem{comment1}The relation $\Delta$({\boldmath $k$}) $= -\Delta$({\boldmath $k+Q_0$}) requires spin singlet pairs.  This is because the exchange interaction of a vector boson gives a repulsive interaction for  singlet pairs; the sign change is crucial in obtaining an attractive interaction.  For triplet pairs, the interaction is already attractive, hence it is necessary that there is no sign change, $\Delta$({\boldmath $k$}) $= \Delta$({\boldmath $k+Q_0$})
\end{thebibliography}
\end{document}